# First Homologous Series of Iron Pnictide Oxide Superconductors $(Fe_2As_2)(Ca_{n+1}(Sc,Ti)_nO_y)$ [$n$ = 3,4,5] with Extremely Thick Blocking Layers

Hiraku Ogino[1,3,a)], Shinya Sato[1,3], Kohji Kishio[1,3] Jun-ichi Shimoyama[1,3], Tetsuya Tohei[2], and Yuichi Ikuhara[2]

[1]Department of Applied Chemistry, The University of Tokyo, 7-3-1 Hongo, Bunkyo-ku, Tokyo 113-8656, Japan

[2]Institute of Engineering Innovation, The University of Tokyo, 2-11-16 Yayoi, Bunkyo-ku, Tokyo 113-8656, Japan

[3]JST-TRIP, Sanban-cho, Chiyoda-ku, Tokyo 102-0075, Japan

a) Electronic mail: tuogino@mail.ecc.u-tokyo.ac.jp

We have discovered first homologous series of iron pnictide oxide superconductors $(Fe_2As_2)(Ca_{n+1}(Sc,Ti)_nO_y)$ [$n$ = 3,4,5]. These compounds have extremely thick blocking layers up to quintuple perovskite oxide layers sandwiched by the $Fe_2As_2$ layers. These samples exhibited bulk superconductivity with relatively high $T_c$ up to 42 K. The relationship between $T_c$ and the iron-plane interlayer distance suggested that superconductivity due to the mono $Fe_2As_2$ layer is substantially 40 K-class.

In a recently discovered layered iron pnictide family[1], the superconducting transition temperature $T_c$ increased under ambient pressure in the order of FeSe ($T_c$ = 8-13 K, $d$ = 5.49 Å)[2,] $(Ba,K)Fe_2As_2$ (38 K, 6.65 Å)[3] and SmFeAs(O,F) (55 K, 8.44 Å)[4], where $d$ represents the distance between the iron-planes. This suggested that emphasis of two-dimensionality by an increase of the interlayer distance of superconducting planes would be a promising method to develop new high-$T_c$ superconductors. On the other hand, a new family of layered iron pnictides with perovskite-type oxide blocking layers has recently been discovered[5-12]. The thickness of the perovskite-type layer can be controlled according to the composition and the synthesis conditions; therefore, an attempt to develop new layered iron pnictides with thicker oxide layers may result in a new series of homologous superconductors. Novel compounds with double or triple perovskite layers were discovered in our recent study, such as $(Fe_2As_2)(Ba_3Sc_2O_5)$ and $(Fe_2As_2)(Sr_4(Sc,Ti)_3O_8)$[12]. These new compounds indicated that a perovskite layer, which can maintain electroneutrality with the $(Fe_2As_2)^{-2}$ layer and have an appropriate lattice size $a$~4 Å, can form layered iron pnictide oxides. According to these empirical observations, three new compounds with different crystal structures, $(Fe_2As_2)(Ca_{n+1}(Sc,Ti)_nO_y)$ [$n$ = 3,4,5 and $y$ ~3$n$-1] were discovered.

Samples were synthesized by solid-state reaction from starting materials of FeAs (3N), Ca (2N), CaO (2N), Ti (3N), $TiO_2$ (3N) and $Sc_2O_3$ (3N). The starting reagents were moisture sensitive, so that manipulation was carried out in a glove box filled with argon gas. Powder mixtures were pelletized, sealed in evacuated quartz ampoules and heated at 1000-1200 °C for 60-100 h followed by slow cooling to room temperature.

Phase identification was carried out using powder XRD (Rigaku Ultima-IV) and intensity data were collected in the $2\theta$ range of 3-80° at a step of 0.02° using Cu-$K_\alpha$ radiation. High resolution TEM images were taken using JEM-2010F and JEM-4010 (JEOL). Magnetic susceptibility was measured using a superconducting quantum interference device (SQUID) magnetometer (Quantum Design MPMS-XL5s). Electrical resistivity was evaluated by the AC four-point-probe method (Quantum Design PPMS).

$(Fe_2As_2)(Ca_4(Sc_{0.67}Ti_{0.33})_3O_y)$, $(Fe_2As_2)(Ca_5(Sc_{0.5}Ti_{0.5})_4O_y)$ and $(Fe_2As_2)(Ca_6(Sc_{0.4}Ti_{0.6})_5O_y)$ samples were successfully obtained by sintering for 72 h at 1050, 1100 and 1200 °C, respectively. Each compound formed as a main phase with small amounts of impurities, such as $CaFe_2As_2$, FeAs and $CaSc_2O_4$, as shown in XRD patterns of Fig. 1. The half unit cell crystal structures of these new compounds are shown in Fig. 2. Their unit cells are tetragonal with a space group of *I4/mmm*. $(Fe_2As_2)(Ca_4(Sc_{0.67}Ti_{0.33})_3O_y)$ has the same crystal structure as $(Fe_2As_2)(Sr_4(Sc,Ti)_3O_8)$[12]. In contrast, the crystal structures of $(Fe_2As_2)(Ca_5(Sc_{0.5}Ti_{0.5})_4O_y)$ and $(Fe_2As_2)(Ca_6(Sc_{0.4}Ti_{0.6})_5O_y)$ have new crystal structure with four and five perovskite layers, respectively, between the $Fe_2As_2$ layers. It should be noted that compounds with larger number of perovskite layers form at higher temperatures. The lattice constants determined from the XRD data were $a$ = 3.922 Å and $c/2$ = 16.78 Å for $(Fe_2As_2)(Ca_4(Sc_{0.67}Ti_{0.33})_3O_y)$, $a$ = 3.902 Å and $c/2$ = 20.63 Å for $(Fe_2As_2)(Ca_5(Sc_{0.5}Ti_{0.5})_4O_y)$, and $a$ = 3.884 Å and $c/2$ = 24.58 Å for $(Fe_2As_2)(Ca_6(Sc_{0.4}Ti_{0.6})_5O_y)$. The half $c$-axis length increased by ~3.9 Å due to a one perovskite layer increase in the half unit cell, which is reasonable, because the lattice constant of cubic $Ca(Sc,Ti)O_3$ is ~3.9 Å. It should be noted that $(Fe_2As_2)(Ca_6(Sc_{0.4}Ti_{0.6})_5O_y)$ has the longest interlayer distance of the iron planes in the layered iron pnictide oxides at ~24.6 Å. On the other hand, small shortenings in the $a$-axis length can be explained by the difference in the ratio of scandium and titanium, because the ionic radius of $Sc^{3+}$ is slightly larger than that of $Ti^{4+}$.

If all the oxygen sites are fully occupied, then the oxygen content $y$ in $(Fe_2As_2)(Ca_4M_3O_y)$, $(Fe_2As_2)(Ca_5M_4O_y)$ and $(Fe_2As_2)(Ca_6M_5O_y)$ is 8, 11 and 14, respectively. However, the samples synthesized from nominal oxygen contents of $y$ = 8, 11 and 14 exhibited very low phase purity with large amounts of $CaTiO_3$ and other phases present. The samples shown in Fig. 1 were synthesized from the oxygen deficient nominal compositions $(Fe_2As_2)(Ca_4(Sc_{0.67}Ti_{0.33})_3O_{7.5})$, $(Fe_2As_2)(Ca_5(Sc_{0.5}Ti_{0.5})_4O_{10})$ and $(Fe_2As_2)(Ca_6(Sc_{0.4}Ti_{0.6})_5O_{13})$. Starting from oxygen deficient compositions with titanium in the lower valence state during the early stage of the reaction might be crucial to form the layered pnictide oxides, because tetravalent titanium tends to form $CaTiO_3$, which is a stable compound. Although the oxygen contents in the resulting samples are difficult to determine, they are considered to be close to 8, 11 or 14, respectively. The partial brown coloring of the quartz ampoule after the reaction suggested that a certain amount of oxygen is supplied to the samples by the partial decomposition of $SiO_2$. The mean valences of the cations at the $M$-site were calculated to be +3.33, +3.50 and +3.60 in $(Fe_2As_2)(Ca_4M_3O_8)$, $(Fe_2As_2)(Ca_5M_4O_{11})$ and $(Fe_2As_2)(Ca_6M_5O_{14})$, respectively, by taking electroneutrality into account. Since the valences of scandium and titanium ions in these samples are considered to be +3 and +4, respectively, the atomic ratio of scandium and titanium is almost uniquely determined in each phase; 2:1 for $(Fe_2As_2)(Ca_4M_3O_8)$, 1:1 for $(Fe_2As_2)(Ca_5M_4O_{11})$ and 2:3 for $(Fe_2As_2)(Ca_6M_5O_{14})$. Samples starting from different ratios of scandium and titanium always contained larger amounts of impurity phases and the variation of lattice constants by changing the scandium/titanium ratio was confirmed to be small. If the scandium/titanium ratio is expressed as $(1\text{-}x):x$, then the relationship between $n$ and $x$ is calculated as $x = 1\text{-}2/n$. Therefore, the general formula of the present system is expressed as $(Fe_2As_2)(Ca_{n+1}(Sc_{2/n}Ti_{1\text{-}2/n})_nO_{3n\text{-}1})$. Further increase in the thickness of the blocking layer, *i.e.*, an increase in $n$, would be possible by tuning the scandium/titanium ratio according to the general formula and synthesis at higher temperatures.

Figure 3 shows high resolution transmission electron microscopy (TEM) images and electron diffraction (ED) patterns taken from the [100] direction of $(Fe_2As_2)(Ca_4(Sc_{0.67}Ti_{0.33})_3O_y)$ and $(Fe_2As_2)(Ca_5(Sc_{0.5}Ti_{0.5})_4O_y)$ crystals. Both TEM images and ED patterns indicate a tetragonal cell with *c*/*a* of ~4.1 for $(Fe_2As_2)(Ca_4(Sc_{0.67}Ti_{0.33})_3O_y)$ and ~5.2 for $(Fe_2As_2)(Ca_5(Sc_{0.5}Ti_{0.5})_4O_y)$. These values correspond well with the values estimated from the XRD analysis; *c*/*a* ~4.3 and 5.3. No stacking faults or superstructures were found along the *c*-axis direction in either crystals. On the other hand, the ED pattern suggested the existence of a superstructure along the *a*-axis in both crystals with a period of 2*a*. An ordered configuration of scandium and titanium in the crystal is not the origin of the superstructure, because the superstructure was found in $(Fe_2As_2)(Ca_4(Sc_{0.67}Ti_{0.33})_3O_y)$ where the ratio of scandium to titanium is 2:1. In addition, such satellite spots were not observed in the ED pattern of $(Fe_2As_2)(Sr_4(Mg_{0.5}Ti_{0.5})_2O_6)$[12]. Therefore, the observed superstructure is probably due to local periodic lattice deformations.

The temperature dependence of the zero-field-cooled (ZFC) and field-cooled (FC) magnetization curves of $(Fe_2As_2)(Ca_4(Sc_{0.67}Ti_{0.33})_3O_y)$, $(Fe_2As_2)(Ca_5(Sc_{0.5}Ti_{0.5})_4O_y)$ and $(Fe_2As_2)(Ca_6(Sc_{0.4}Ti_{0.6})_5O_y)$ measured under 1 Oe are shown in Fig. 4(a). All of the compounds exhibited large diamagnetism, which suggests superconductivity. The $T_{c(onset)}$'s were ~36 K for $(Fe_2As_2)(Ca_5(Sc_{0.5}Ti_{0.5})_4O_y)$ and $(Fe_2As_2)(Ca_6(Sc_{0.4}Ti_{0.6})_5O_y)$ and 31 K for $(Fe_2As_2)(Ca_4(Sc_{0.67}Ti_{0.33})_3O_y)$. The superconducting volume fraction estimated from the ZFC magnetization at 2 K was much larger than the perfect diamagnetism due to the demagnetization effect and the porous microstructure with closed pores. The reversible magnetization observed down to several K below the $T_{c(onset)}$ suggested poor grain coupling and weak intragrain pinning at high temperatures.

Figure 4(b) shows the temperature dependence of resistivity for

$(Fe_2As_2)(Ca_4(Sc_{0.67}Ti_{0.33})_3O_y)$, $(Fe_2As_2)(Ca_5(Sc_{0.5}Ti_{0.5})_4O_y)$ and $(Fe_2As_2)(Ca_6(Sc_{0.4}Ti_{0.6})_5O_y)$, with resistivity curves from 25 to 50 K in the inset. All three samples exhibited metallic behavior in the normal state with respective $T_{c(onset)}$'s of approximately 33, 41 and 42 K. The broad transitions observed for all samples are partly explained by poor grain coupling; however, apparently higher $T_{c(onset)}$'s than those observed in the magnetization measurements may indicate the effect of superconductivity fluctuation near $T_c$.

In the series of newly discovered compounds $(Fe_2As_2)(Ca_{n+1}(Sc,Ti)_nO_y)$ [$n$ = 3,4,5 and $y$ ~3$n$-1], superconducting $Fe_2As_2$ layers are largely separated from each other by the thick perovskite oxide layers. To our knowledge, the interlayer distance between the iron planes, $d$ ~24.6 Å, of $(Fe_2As_2)(Ca_6(Sc_{0.4}Ti_{0.6})_5O_y)$ is the longest among all superconductors, except for artificially deposited multi-layer films. It should be noted that $d$ of layered iron-based superconductors can be changed from ~5 Å for FeSe[2] to 24.6 Å. The variation of $d$ is much larger than other layered superconductors, such as cuprate superconductors, which indicates the high chemical stability and structural flexibility of the $Fe_2As_2$ layer. The $T_{c(onset)}$'s of the iron pnictides and chalcogenides[1,3,4,6,9,10,13-20] are plotted as a function of the interlayer distance $d$ in Fig. 5. In this plot, superconductors having a doped iron plane were excluded, because such compounds usually exhibit lower $T_c$ due to structural and/or electronic disorder in the superconducting layer. No systematic dependence of $T_c$ on $d$ was found, but $T_c$ seems almost constant at ~40 K when $d$ > 15 Å. In other words, the dimensionality of the crystal structure is not a major factor in determining $T_c$ of layered iron-based superconductors. The coherence length in the $c$-axis direction, $\xi_c$, of the 1111 compound is reported to be 4.5 Å[21].

Although $\xi_c$ of the $(Fe_2As_2)(Ca_{n+1}(Sc,Ti)_nO_y)$ series has not been evaluated, it is believed to be shorter than that of the 1111 compound, due to the thick blocking layers. Therefore, superconductivity in $(Fe_2As_2)(Ca_{n+1}(Sc,Ti)_nO_y)$ compounds is considered to originate from completely separated $Fe_2As_2$ layers, which inherently exhibit 40 K-class $T_c$. The particularly high $T_c$ of the 1111 compounds may be due to moderate interaction between the $Fe_2As_2$ layers.

New iron pnictide oxide superconductors with extremely thick blocking layers $(Fe_2As_2)(Ca_{n+1}(Sc,Ti)_nO_y)$ [$n$ = 3,4,5 and $y$ ~3$n$-1] were discovered in the present study. These samples exhibited bulk superconductivity with relatively high $T_c$ up to 42 K. The relationship between $T_c$ and the iron-plane interlayer distance suggested that superconductivity due to the mono $Fe_2As_2$ layer is substantially 40 K-class. It is thought that there is still considerable room for the development of new layered iron pnictide compounds, due to the structural and chemical flexibility of the blocking layer. However, the present results suggest that optimization of the local structure of $Fe_2As_2$ layers and the dimensionality of the crystal structure may not lead to further enhancement of $T_c$. Other challenging approaches, such as the introduction of double superconducting layers, will be required for a new breakthrough in $T_c$.

This work was supported in part by the Ministry of Education, Culture, Sports, Science, and Technology (MEXT), Japan, through a Grant-in-Aid for Young Scientists (B) (No. 21750187).

**Figure captions**

Figure 1. Powder XRD patterns of (a) $(Fe_2As_2)(Ca_4(Sc_{0.67}Ti_{0.33})_3O_y)$, (b) $(Fe_2As_2)(Ca_5(Sc_{0.5}Ti_{0.5})_4O_y)$, and (c) $(Fe_2As_2)(Ca_6(Sc_{0.4}Ti_{0.6})_5O_y)$.

Figure 2. Crystal structures of (a) $(Fe_2As_2)(Ca_4(Sc,Ti)_3O_8)$, (b) $(Fe_2As_2)(Ca_5(Sc,Ti)_4O_{11})$, and (c) $(Fe_2As_2)(Ca_6(Sc,Ti)_5O_{14})$.

Figure 3. High resolution TEM images and corresponding ED patterns of (a) $(Fe_2As_2)(Ca_4(Sc_{0.67}Ti_{0.33})_3O_y)$ and (b) $(Fe_2As_2)(Ca_5(Sc_{0.5}Ti_{0.5})_4O_y)$ crystals viewed from the [100] direction.

Figure 4. Superconducting properties. (a) Temperature dependence of ZFC and FC magnetization curves for $(Fe_2As_2)(Ca_4(Sc_{0.67}Ti_{0.33})_3O_y)$, $(Fe_2As_2)(Ca_5(Sc_{0.5}Ti_{0.5})_4O_y)$ and $(Fe_2As_2)(Ca_6(Sc_{0.4}Ti_{0.6})_5O_y)$ bulk samples measured under 1 Oe. (b) Temperature dependence of resistivity for bulk $(Fe_2As_2)(Ca_4(Sc_{0.67}Ti_{0.33})_3O_y)$, $(Fe_2As_2)(Ca_5(Sc_{0.5}Ti_{0.5})_4O_y)$ and $(Fe_2As_2)(Ca_6(Sc_{0.4}Ti_{0.6})_5O_y)$. Magnified resistivity curves are shown in the inset.

Figure 5. Relationship between $T_c$ and the iron-plane interlayer distance for various layered iron pnictides.

Fig. 1

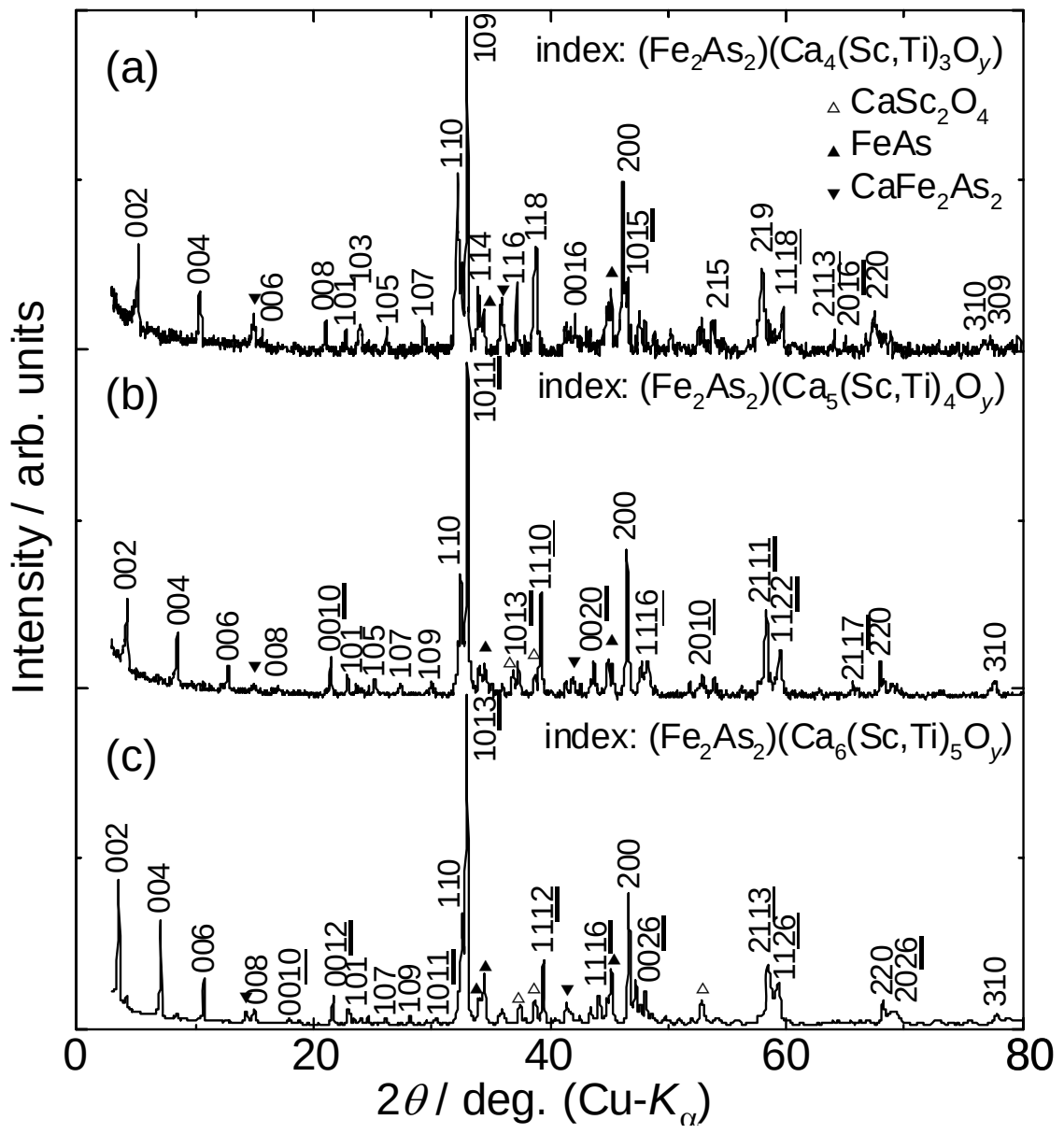

Fig. 2

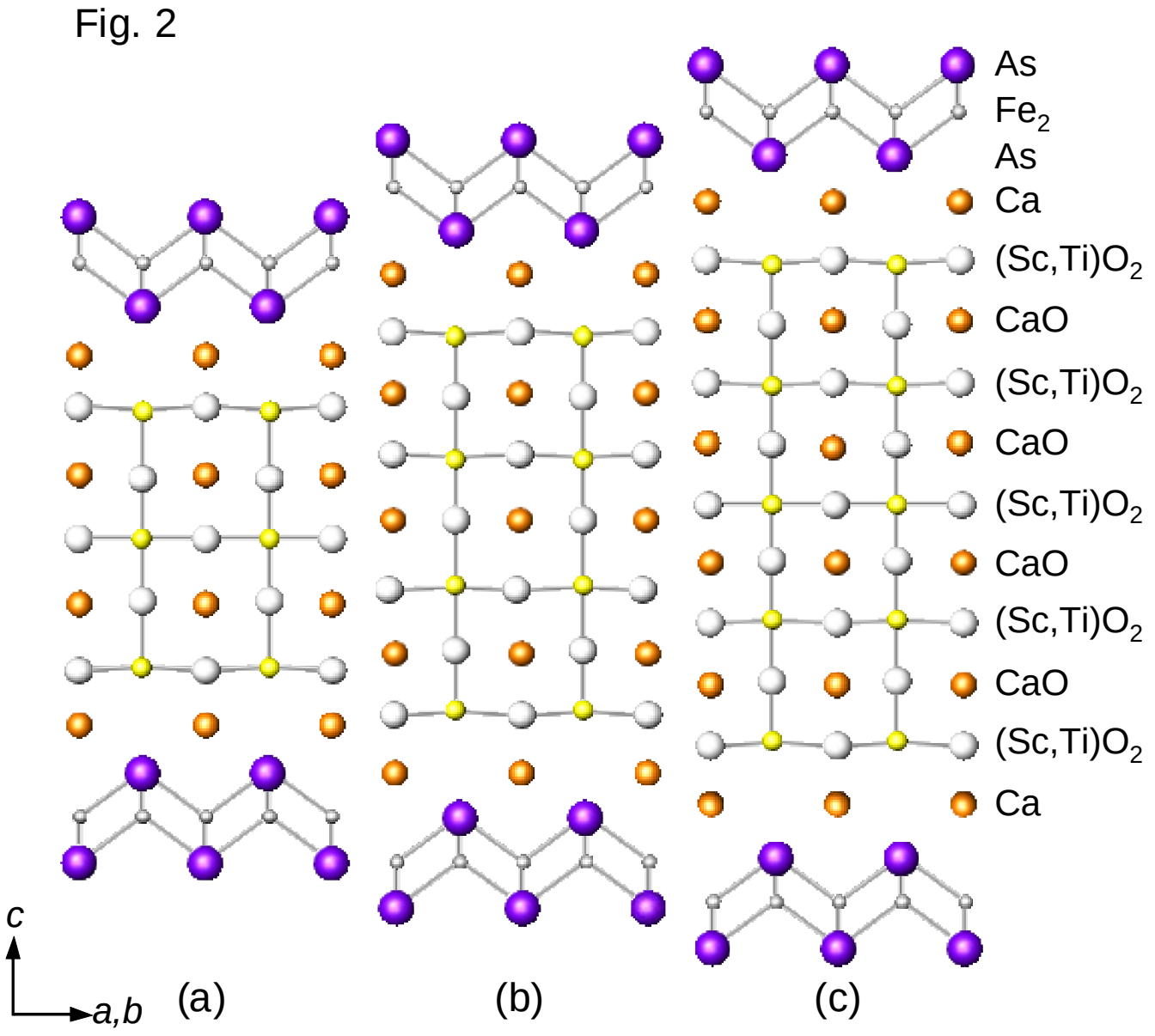

Fig. 3

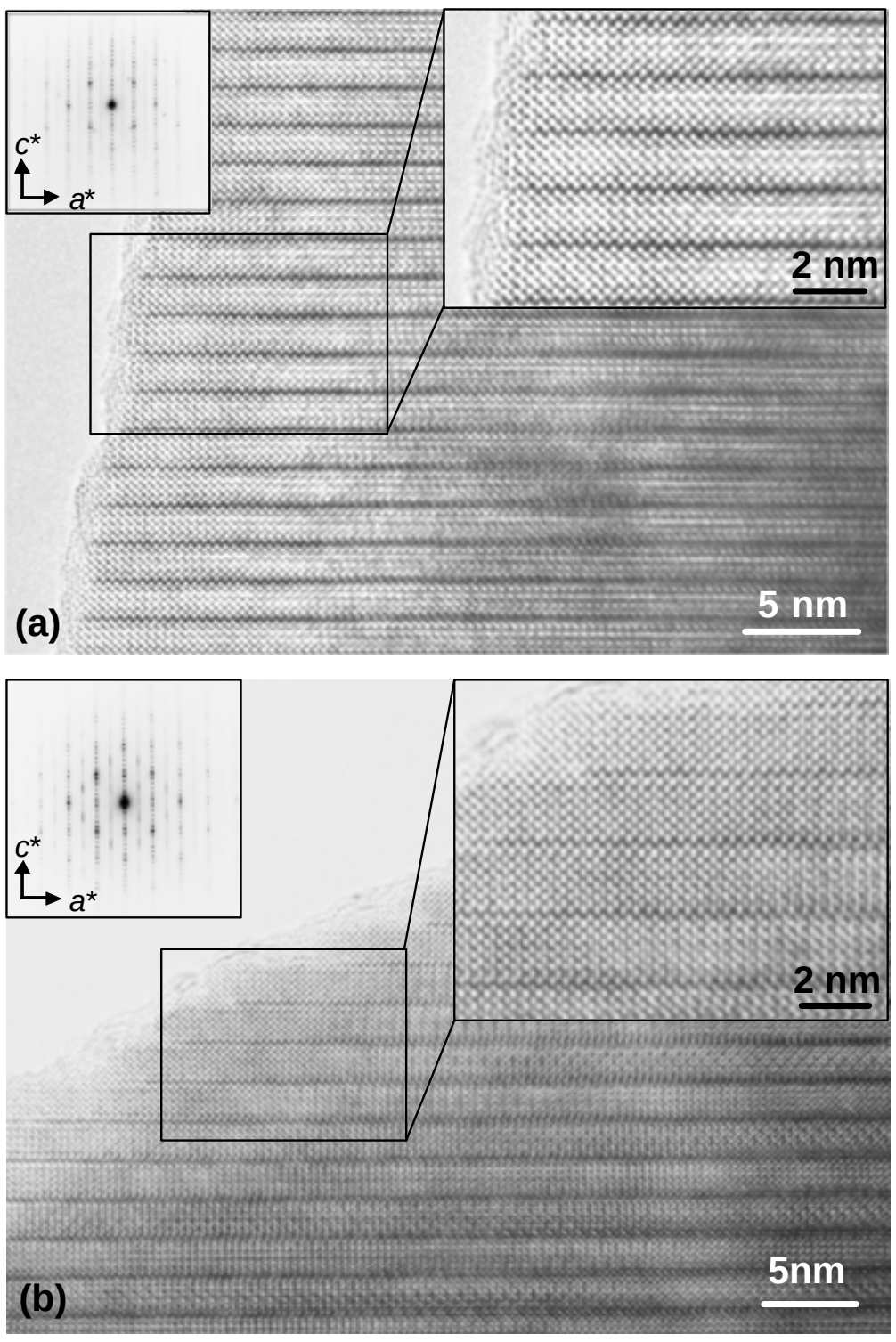

Fig. 4

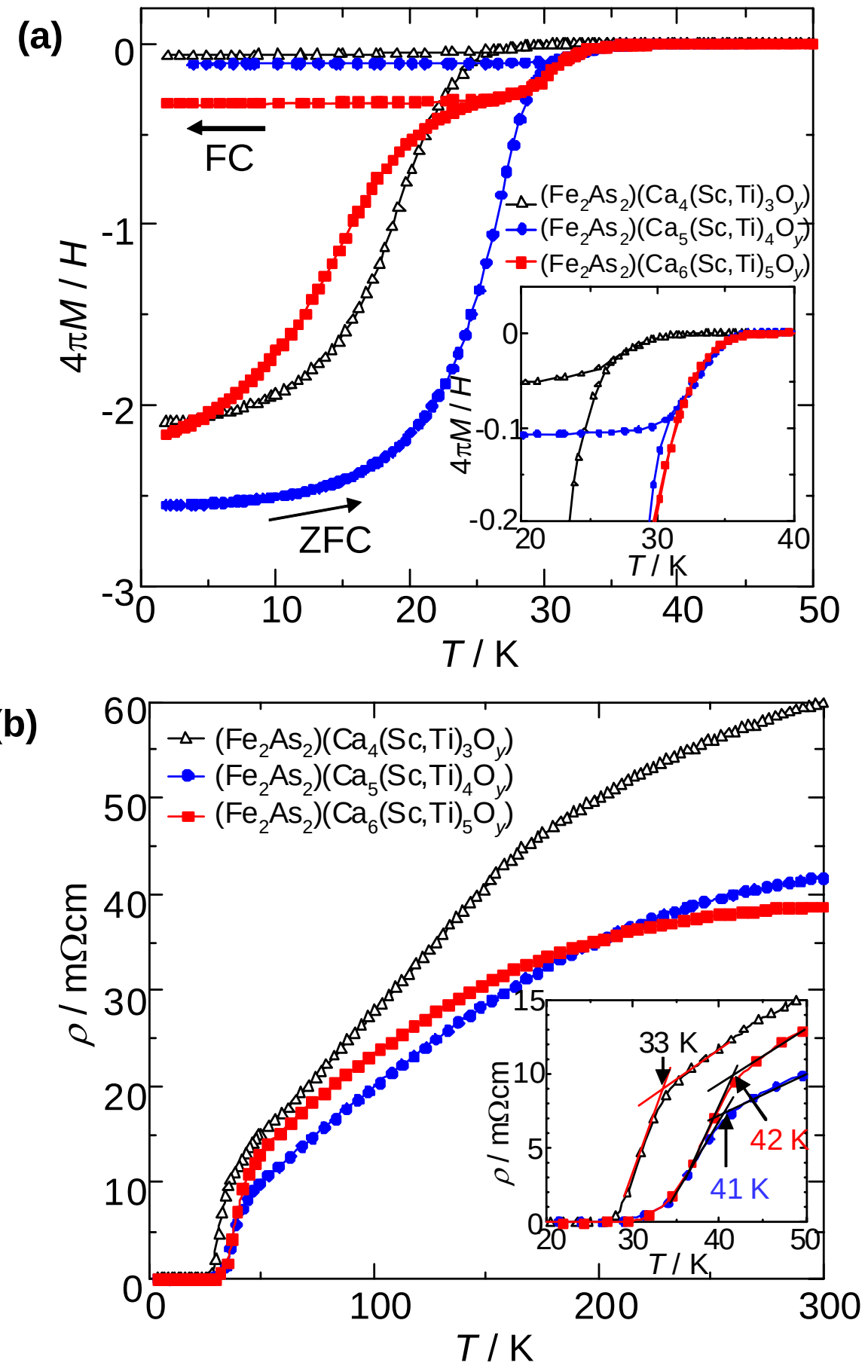

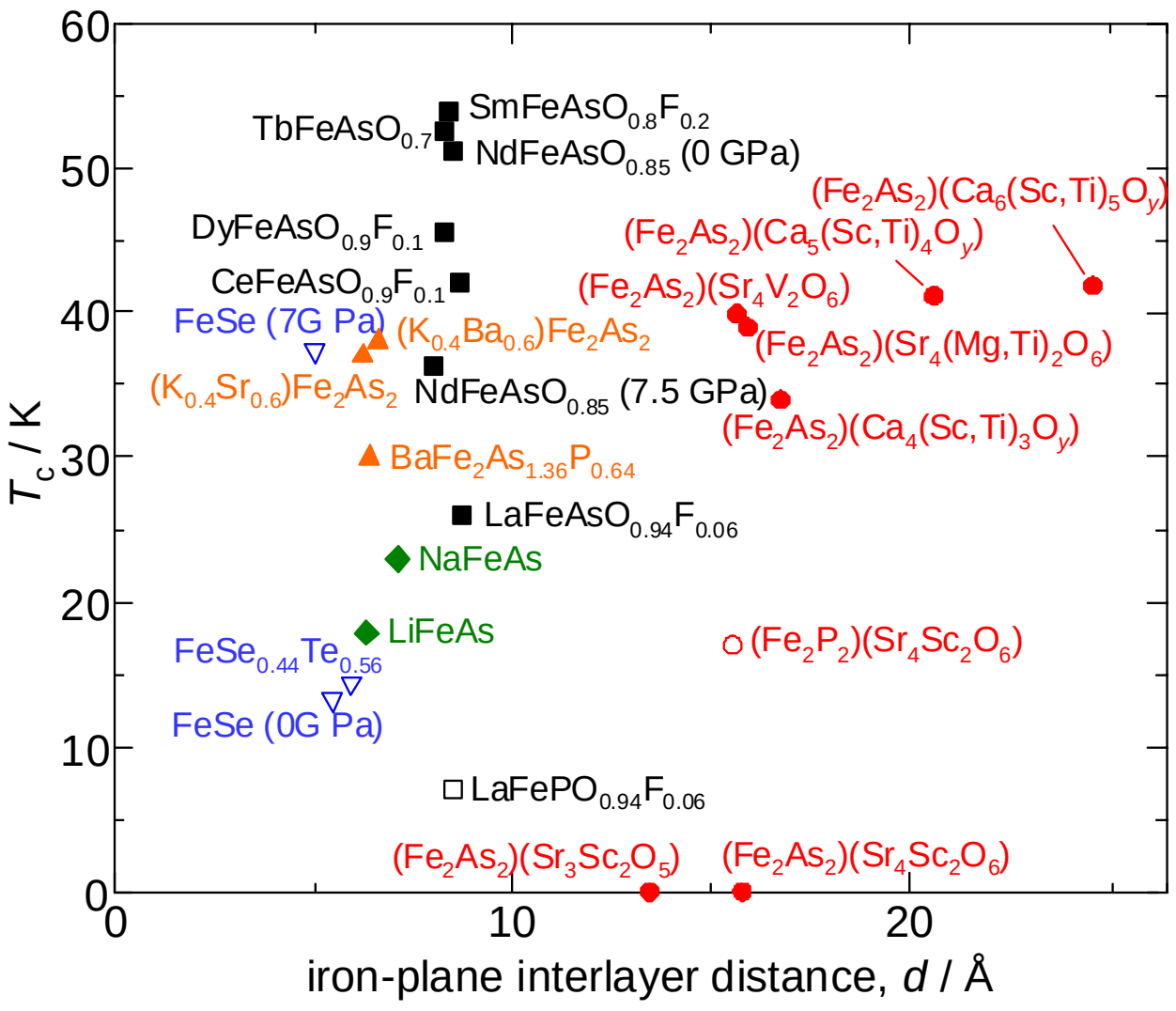